\documentclass[11pt]{article}
\usepackage{amsfonts}
\def \sech{\mathop{\rm sech}\nolimits}
\def \csch{\mathop{\rm csch}\nolimits}
\usepackage{amssymb,amsmath,mathrsfs}
\usepackage{graphicx}
\usepackage{subfigure}

\begin{document}
\title{\bf{Nonlocal nonlinear Schr\"{o}dinger equation and its discrete version: soliton solutions and gauge equivalence} }
\author{ Li-Yuan Ma, Zuo-Nong Zhu
\footnote{Corresponding author. Email: znzhu@sjtu.edu.cn}
\\
Department of Mathematics, Shanghai Jiao Tong University,\\ 800 Dongchuan Road, Shanghai, 200240, P. R. China\\
}
\date{March 23,2015}
\maketitle

\begin{abstract}
 In this paper, we try to understand the geometry for a nonlocal nonlinear Schr\"{o}dinger equation (nonlocal NLS) and its discrete version introduced by Ablowitz and Musslimani.
 We show that, under the gauge transformations, the nonlocal focusing NLS and the nonlocal defocusing NLS are, respectively, gauge equivalent to a Heisenberg-like equation and a modified Heisenberg-like equation, and their discrete versions are, respectively, gauge equivalent to a discrete Heisenberg-like equation and a discrete modified Heisenberg-like equation. From the gauge equivalence, although the geometry related to the nonlocal NLS is not very clear, we can see that the properties between the nonlocal NLS and its discrete version and NLS and discrete NLS have big difference. By constructing the Darboux transformation for discrete nonlocal NLS equations including the cases of focusing and defocusing, we derive their discrete soliton solutions, which differ from the ones obtained by using the scattering transformation.
\end{abstract}

PACS number(s): 05.45.Yv, 02.30.Ik\\
\section{Introduction}
Very recently, Ablowitz and Musslimani investigated a nonlocal NLS equation \cite{1}
\begin{equation}\label{nls}
i q_t(x,t)+q_{xx}(x,t)\pm 2q(x,t)q^{*}(-x,t)q(x,t)=0,
\end{equation}
which is derived from a new symmetry reduction of the well-known AKNS system,
where $q(x,t)$ is a complex valued function of the real variables $x$ and $t$ and $*$ denotes complex conjugation.
The nonlocal NLS equation \eqref {nls} is a new integrable system possessing the Lax pair, infinitely many conservation laws and it is solvable by using the inverse scattering transformation.
Like the classical NLS equation,
the nonlocal equation \eqref{nls} keeps the following parity-time transform invariant
\begin{equation}
x\rightarrow -x, \quad t\rightarrow -t, \quad q(x,t)\rightarrow q^{*}(x,t).
\end{equation}
Hence, it is $PT$ symmetric \cite{2} and can be regarded as a mathematical model describing wave propagation phenomena
in $PT$ symmetric nonlinear media \cite{3,4,5}. The nonlocal NLS has attracted the attention of researchers since its special properties. For example, by using the inverse scattering transformation, Ablowitz and Musslimani obtained its breather solution \cite{1}; In \cite{6} Sarma etal showed that
the $PT$-symmetric nonlocal NLS \eqref{nls} can simultaneously support both bright and dark soliton solutions;
Valchev \cite{7} studied some basic properties of the nonlocal focusing NLS  equation including its scattering operator,
the special solutions by using the dressing method, and the Hamiltonian formalism. In addition, dark and antidark soliton interactions in the nonlocal defocusing NLS has been discussed \cite {8}. \\
On the other hand, Ablowitz and Musslimani \cite{9} also investigated a discrete version of the nonlocal NLS \eqref{nls}
\begin{eqnarray}\label{dnls}
i \frac{d Q_n}{d\tau}+Q_{n+1}+Q_{n-1}-2Q_n\pm Q_n Q^{*}_{-n}(Q_{n+1}+Q_{n-1})=0,
\end{eqnarray}
which is a discrete $PT$ symmetric model, and it contains a linear Lax pair and an infinite many conservation laws. In Ref. 9,
a discrete one-soliton solutions with the unique features of power oscillations and singularity formation has been derived by using a left-right Riemann-Hilbert formulation. In \cite{6} Sarma etal investigated another discrete $PT$ symmetric nonlocal NLS
\begin{eqnarray}
i \frac{d a_n}{d\tau}+k(a_{n+1}+a_{n-1})+a_n^2 a^{*}_{-n}=0.
\end{eqnarray}
We have known that the focusing NLS and defocusing NLS are, respectively, gauge equivalent to the Schr\"odinger flow of maps from $R^1$ into $S^2$ in $R^3$ and from $R^1$ into $H^2$ in $R^{2+1}$ \cite{10}-\cite{14}. This gives the geometric explanations for the NLS equation. The geometry related to the discrete NLS has also been discussed \cite{15,16}. So, what is
geometric interpretation for the nonlocal NLS and its discrete version?
In this paper, we try to understand the geometry for the nonlocal NLS \eqref{nls} and its discrete version \eqref{dnls}. We will show that, under the gauge transformations, the nonlocal focusing NLS and the nonlocal defocusing NLS are, respectively, gauge equivalent to a Heisenberg-like equation and a modified Heisenberg-like equation, and their discrete versions are, respectively, gauge equivalent to a discrete Heisenberg-like equation and a discrete modified Heisenberg-like equation. From the gauge equivalence, although the geometry related to the nonlocal NLS and its discrete version is not very clear, we can see that the properties between the nonlocal NLS and its discrete version and NLS and discrete NLS have big differences. By constructing the Darboux transformation for discrete nonlocal NLS equations including two cases of focusing and defocusing, we also derive their discrete soliton solutions, which differ from the ones obtained by using the scattering transformation \cite{9}.\\

\section{Gauge equivalent structures of the nonlocal focusing NLS and the nonlocal defocusing NLS}

In this section, we try to understand the geometry related to the nonlocal focusing NLS and the nonlocal defocuing NLS through investigating their gauge equivalent structures. We will show that the nonlocal focusing NLS and the nonlocal defocuing NLS are, respectively, gauge equivalent to a Heisenberg-like equation and a modified Heisenberg-like equation. From the gauge equivalence, we can see that there exist big differences between the nonlocal NLS equation and NLS equation.\\
We first recall that the nonlocal focusing NLS equation
\begin{eqnarray}\label{NLS1}
i q_t(x,t)+q_{xx}(x,t)+ 2q(x,t)q^{*}(-x,t)q(x,t)=0,
\end{eqnarray}
is yielded by the integrability condition of the following linear problem \cite{1}:
\begin{equation}\label{lp}
\varphi_x=M \varphi,\quad \varphi_t=N \varphi
\end{equation}
with
\begin{align*}
M=\left(
\begin{array}{cc}
-i\lambda & q(x,t) \\
-q^{*}(-x,t) & i\lambda  \\
\end{array}\right),\quad
N=\left(
\begin{array}{cc}
-2i\lambda^2+iq(x,t)q^{*}(-x,t) & 2\lambda q(x,t)+iq_x(x,t) \\
-2\lambda q^{*}(-x,t)+iq^{*}_x(-x,t) & 2i\lambda^2-iq(x,t)q^{*}(-x,t)  \\
\end{array}\right).
\end{align*}
Under the following gauge transformation \cite{11}:
\begin{equation}
\tilde{M}=G^{-1}M G-G^{-1}G_{x}, \quad \tilde{N}=G^{-1}N G-G^{-1}G_{t},
\end{equation}
where $G$ is a solution of the system \eqref{lp} for $\lambda=0$, i.e.,
\begin{equation}\label{gt}
G_{x}=M(0)G, \quad G_{t}=N(0)G,
\end{equation}
we can obtain
\begin{align}\label{wave}
\tilde{M}=-i\lambda G^{-1}\sigma_3 G \triangleq -i\lambda S, \quad
\tilde{N}=-2i\lambda^2S+\lambda SS_x.
\end{align}
The compatibility condition $\tilde{M}_t-\tilde{N}_x+[\tilde{M},\tilde{N}]=0$ yields a Heisenberg-like equation
\begin{eqnarray}\label{HE}
S_t=\frac{i}{2}[S, S_{xx}].
\end{eqnarray}
From the structure of the matrix $M(0)$ and $N(0)$, we can see that $G$ in \eqref{gt} has the form
\begin{equation*}
G=\left(
\begin{array}{cc}
f(x,t) & g^{*}(-x,t) \\
g(x,t) & f^{*}(-x,t) \\
\end{array}\right).
\end{equation*}
Hence the structure of the matrix $S$ in equation \eqref{HE} can be given by
\begin{equation}\label{s}
S=G^{-1}\sigma_3 G=\frac{1}{\Omega}\left(
\begin{array}{cc}
f(x,t)f^{*}(-x,t)+g(x,t)g^{*}(-x,t) & 2f^{*}(-x,t)g^{*}(-x,t) \\
-2f(x,t)g(x,t) & -f(x,t)f^{*}(-x,t)-g(x,t)g^{*}(-x,t) \\
\end{array}\right),
\end{equation}
where $\Omega=f(x,t)f^{*}(-x,t)-g(x,t)g^{*}(-x,t)$. We should remark here that we say equation \eqref{HE} is a Heisenberg-like equation since it has the same form as the Heisenberg equation, but there exists a big difference between \eqref{HE} and the Heisenberg equation. In fact, in the case of focusing NLS, i.e., $q^*(-x,t)=q^*(x,t)$, $G$ in equation \eqref{gt} has the form
\begin{eqnarray*}
G=\left(\begin{array}{cc}
f(x,t) & -g^{*}(x,t) \\
g(x,t) & f^{*}(x,t) \\
\end{array}\right),
\end{eqnarray*}
and thus $S=G^{-1}\sigma_3 G$ has the form
\begin{eqnarray*}
S=\frac{1}{|f|^2+|g|^2}\left(
\begin{array}{cc}
|f|^2-|g|^2& -2f^{*}g^{*}\\
-2fg & |g|^2-|f|^2 \\
\end{array}\right).
\end{eqnarray*}
Set $f=a(x,t)+ib(x,t), g=c(x,t)+id(x,t)$, then the matrix $S$ can be written as
\begin{eqnarray*}
S=\left(
\begin{array}{cc}
s_{1}& s_{2}-is_{3}\\
 s_{2}+is_{3} &  -s_{1}\\
\end{array}\right),
\end{eqnarray*}
where the vector $\textbf{S}$=$(s_{1},s_{2},s_{3})^T\in S^2$ in $R^3$, and $s_{j}, j=1,2,3$ is given by
\begin{eqnarray*}
s_{1}=\frac{|f|^2-|g|^2}{|f|^2+|g|^2},\qquad
s_{2}=\frac{2(bd-ac)}{|f|^2+|g|^2},\qquad
s_{3}=\frac{-2(bc+ad)}{|f|^2+|g|^2}.
\end{eqnarray*}
So, equation \eqref{HE} reduces to the Heisenberg equation
\begin{eqnarray}
\textbf{S}_t=\textbf{S}\times \textbf{S}_{xx}.
\end{eqnarray}
For the one-soliton solution of the nonlocal focusing NLS \eqref{NLS1}
\begin{align}\label{one}
q(x,t)=2iae^{2bx-4i(a^2-b^2)t}\csch(8abt-2iax)),
\end{align}
we solve equation  \eqref{gt} as follows:
\begin{eqnarray*}
G=\left(
\begin{array}{cc}
-\frac{ia e^{2bx-4i(a^2-b^2)t}\csch(8abt-2iax)}{\sqrt{a^2+b^2}} & \frac{-b+i a\coth(8abt-2iax)}{\sqrt{a^2+b^2}} \\
\frac{-b-i a\coth(8abt-2iax)}{\sqrt{a^2+b^2}} & \frac{i ae^{-2bx+4i(a^2-b^2)t}\csch(8abt-2iax)}{\sqrt{a^2+b^2}} \\
\end{array}\right).
\end{eqnarray*}
 Hence $S$ can be solved as
\begin{equation*}
S\!=\!\!\!\left(\!\!\!
\begin{array}{cc}
1+\frac{2a^2\csch^2(8abt-2iax)}{a^2+b^2} & \frac{-2a e^{-2bx +4i(a^2-b^2)t}(ib+a\coth(8abt-2iax))\csch(8abt-2iax)}{a^2+b^2} \\
\frac{2a e^{2bx -4i(a^2-b^2)t}(-ib+a\coth(8abt-2iax))\csch(8abt-2iax)}{a^2+b^2} & -1-\frac{2a^2\csch^2(8abt-2iax)}{a^2+b^2} \\
\end{array}\!\!\!\right).
\end{equation*}
Set $f=a(x,t)+ib(x,t), g(x,t)=c(x,t)+id(x,t)$, then the matrix $S$ in equation \eqref{s} can be written as
\begin{equation*}
S=\left(
\begin{array}{cc}
s_1(x,t)+i s_2(x,t) & -(s_3(-x,t)-i s_4(-x,t)) \\
s_3(x,t)+i s_4(x,t) & -(s_1(x,t)+i s_2(x,t)) \\
\end{array}\right),
\end{equation*}
where
\begin{eqnarray}\label{c1}
&&s_1(x,t)s_1(-x,t)+s_2(x,t)s_2(-x,t)-s_3(x,t)s_3(-x,t)-s_4(x,t)s_4(-x,t)=1,\nonumber\\
&&2s_1(x,t)s_2(x,t)+s_3(x,t)s_4(-x,t)-s_4(x,t)s_3(-x,t)=0,
\end{eqnarray}
 and $s_j (j=1,2,3,4)$ is given by
\begin{equation*}
s_1(x,t)=\frac{s_{11}}{\Gamma},\quad s_2(x,t)=\frac{s_{12}}{\Gamma},
\quad s_3(x,t)=\frac{s_{21}}{\Gamma},\quad s_4(x,t)=\frac{s_{22}}{\Gamma},
\end{equation*}
with
\begin{align}\nonumber
\Gamma=&(a^2(x,t)+b^2(x,t))(a^2(-x,t)+b^2(-x,t))+(c^2(x,t)+d^2(x,t))(c^2(-x,t)+d^2(-x,t))\\\nonumber
&-2(a(x,t)c(x,t)+b(x,t)d(x,t))(a(-x,t)c(-x,t)+b(-x,t)d(-x,t))\\\nonumber
&-2(b(x,t)c(x,t)-a(x,t)d(x,t))(b(-x,t)c(-x,t)-a(-x,t)d(-x,t)),\\\nonumber
s_{11}=&(a^2(x,t)+b^2(x,t))(a^2(-x,t)+b^2(-x,t))-(c^2(x,t)+d^2(x,t))(c^2(-x,t)+d^2(-x,t)),\\\nonumber
s_{12}=&2(a(x,t)c(x,t)+b(x,t)d(x,t))(b(-x,t)c(-x,t)-a(-x,t)d(-x,t))\\\nonumber
&-2(b(x,t)c(x,t)-a(x,t)d(x,t))(b(-x,t)d(-x,t)+a(-x,t)c(-x,t)),\\\nonumber
s_{21}=&2(a^2(x,t)+b^2(x,t))(d(x,t)b(-x,t)-c(x,t)a(-x,t))\\\nonumber
&+2(c^2(x,t)+d^2(x,t))(a(x,t)c(-x,t)-b(x,t)d(-x,t)),\\\nonumber
s_{22}=&-2(a^2(x,t)+b^2(x,t))(d(x,t)a(-x,t)+c(x,t)b(-x,t))\\\nonumber
&+2(c^2(x,t)+d^2(x,t))(b(x,t)c(-x,t)+a(x,t)d(-x,t)).
\end{align}
We thus see that equation \eqref{HE} can be transformed into
\begin{align}
\frac{ds_1}{dt}=&-\frac{1}{2}\left(s_4s_{3xx}(-x)-s_3(-x)s_{4xx}+s_{3xx}s_4(-x)-s_3s_{4xx}(-x) \right),\nonumber\\
\frac{ds_2}{dt}=&\frac{1}{2}\left(s_3s_{3xx}(-x)-s_3(-x)s_{3xx}+s_4s_{4xx}(-x)-s_4(-x)s_{4xx}\right),\nonumber\\
\frac{ds_3}{dt}=&s_2s_{3xx}-s_3s_{2xx}+s_1s_{4xx}-s_4s_{1xx},\nonumber\\
\frac{ds_4}{dt}=&s_3s_{1xx}-s_1s_{3xx}+s_2s_{4xx}-s_4s_{2xx}.
\end{align}
Next we will discuss the gauge equivalence for the nonlocal defocusing NLS equation
\begin{eqnarray}\label{NLS2}
i q_t(x,t)+q_{xx}(x,t)- 2q(x,t)q^{*}(-x,t)q(x,t)=0,
\end{eqnarray}
 which has the Lax pair \cite{1}
 \begin{equation}
\varphi_x=M \varphi,\quad \varphi_t=N \varphi
\end{equation}
where
\begin{align*}
M=\left(
\begin{array}{cc}
\lambda & q^{*}(-x,t) \\
 q(x,t) & -\lambda  \\
\end{array}\right),\quad
N=i \left(
\begin{array}{cc}
-2\lambda^2+ q(x,t)q^{*}(-x,t) & -2\lambda q^{*}(-x,t)+q^{*}_x(-x,t) \\
- 2\lambda q(x,t) + q_x(x,t) & 2\lambda^2- q(x,t)q^{*}(-x,t)  \\
\end{array}\right).
\end{align*}
Under gauge transformation:
\begin{equation}
\tilde{M}=G^{-1}M G-G^{-1}G_{x}, \quad \tilde{N}=G^{-1}N G-G^{-1}G_{t},
\end{equation}
where $G$ satisfies
\begin{equation}\label{mgt}
G_{x}=M(0)G, \quad G_{t}=N(0)G,
\end{equation}
we can obtain
\begin{align}
\tilde{M}=\lambda G^{-1}\sigma_3 G \triangleq i\lambda S, \quad
\tilde{N}=2\lambda^2S+i \lambda SS_x,
\end{align}
with $S$ being defined by $S=-iG^{-1}\sigma_3 G$.
The compatibility condition $\tilde{M}_t-\tilde{N}_x+[\tilde{M},\tilde{N}]=0$ yields a modified Heisenberg-like equation
\begin{eqnarray}\label{mHE}
S_t=\frac{1}{2}[S,S_{xx}].
\end{eqnarray}
From the structure of the matrix $M(0)$ and $N(0)$, we can see that $G$ in \eqref{mgt} has the form
\begin{equation*}
G=\left(
\begin{array}{cc}
f(x,t) & -g^{*}(-x,t) \\
g(x,t) & f^{*}(-x,t) \\
\end{array}\right).
\end{equation*}
Hence the matrix $S$ is given by
\begin{equation}\label{s2}
S=\frac{i}{\Omega}\left(
\begin{array}{cc}
g(x,t)g^{*}(-x,t)-f(x,t)f^{*}(-x,t) & 2f^{*}(-x,t)g^{*}(-x,t) \\
2f(x,t)g(x,t) & f(x,t)f^{*}(-x,t)-g(x,t)g^{*}(-x,t) \\
\end{array}\right).
\end{equation}
where $\Omega=f(x,t)f^{*}(-x,t)+g(x,t)g^{*}(-x,t)$. We remark here that we say equation \eqref{mHE} is a modified Heisenberg-like equation since it has the same form as the modified Heisenberg equation, but it has big difference with the modified Heisenberg equation. In fact, we see that in the case of defocusing NLS, the matrix $G$ in equation \eqref{mgt} has the form
\begin{eqnarray*}
G=\left(\begin{array}{cc}
f(x,t) & g(x,t) \\
f^*(x,t) & -g^{*}(x,t) \\
\end{array}\right),
\end{eqnarray*}
and thus $S=-iG^{-1}\sigma_3 G$ has the form
\begin{eqnarray*}
S=\frac{i}{Re(fg^*)}\left(
\begin{array}{cc}
-i Im(fg^*)& -|g|^2\\
-|f|^2 & i Im(fg^*)\\
\end{array}\right).
\end{eqnarray*}
Set $f=a(x,t)+ib(x,t), g=c(x,t)+id(x,t)$, then the matrix $S$ can be rewritten as
\begin{eqnarray*}
S=\left(
\begin{array}{cc}
s_{1}& i(s_{3}-s_{2})\\
 i(s_{2}+s_{3}) &  -s_{1}\\
\end{array}\right),
\end{eqnarray*}
where the vector $\textbf{S}$=$(s_{1},s_{2},s_{3})^T\in H^2$ in $R^{2+1}$, i.e., $s_1^2+s_2^2-s_3^2=-1,$ and $s_{j}, j=1,2,3$ is given by
\begin{eqnarray*}
s_{1}=\frac{bc-ad}{ac+bd},\qquad
s_{2}=\frac{|g|^2-|f|^2}{2(ac+bd)},\qquad
s_{3}=\frac{-|f|^2-|g|^2}{2(ac+bd)}.
\end{eqnarray*}
Thus, the modified Heisenberg-like equation \eqref{mHE} leads to the modified Heisenberg equation
\begin{eqnarray}
\textbf{S}_t=\textbf{S}\dot{\times} \textbf{S}_{xx},
\end{eqnarray}
where  $\dot{\times}$
denotes the pseudo cross product in $R^{2+1}$ defined by
$\textbf{a}\dot{\times}\textbf{b}=
(a_2b_3-a_3b_2,a_3b_1-a_1b_3,-(a_1b_2-a_2b_1) )$.\\
For a solution $q(x,t)$ of nonlocal defocusing NLS given by
\begin{align}\label{fuone}
q(x,t)=-2ibe^{-2ax+4i(a^2-b^2)t}\sech(8abt+2ibx)),
\end{align}
the solution to equation \eqref{mgt} is
\begin{equation*}
G=\left(
\begin{array}{cc}
 \frac{-a-i b\tanh(8abt+2ibx)}{\sqrt{a^2+b^2}}& \frac{b e^{2ax-4i(a^2-b^2)t}\sech(8abt+2ibx)}{\sqrt{a^2+b^2}}  \\
\frac{-b e^{-2ax+4i(a^2-b^2)t}\sech(8abt+2ibx)}{\sqrt{a^2+b^2}} & \frac{-a+i b\tanh(8abt+2ibx)}{\sqrt{a^2+b^2}} \\
\end{array}\right),
\end{equation*}
and
\begin{equation*}
S=\left(
\begin{array}{cc}
 1-\frac{2b^2\sech^2(8abt+2ibx)}{a^2+b^2}& -\frac{2b e^{-4i(a^2-b^2)t+2ax}\sech(8abt+2ibx)\left(a-ib \tanh(8abt+2ibx)\right)}{a^2+b^2}  \\
-\frac{2b e^{4i(a^2-b^2)t-2ax}\sech(8abt+2ibx)\left(a+ib \tanh(8abt+2ibx)\right)}{a^2+b^2} & -1+\frac{2b^2\sech^2(8abt+2ibx)}{a^2+b^2} \\
\end{array}\right).
\end{equation*}
Set $f=a(x,t)+ib(x,t), g(x,t)=c(x,t)+id(x,t)$, then the matrix $S$ in equation \eqref{s2} can be written as
\begin{equation*}
S=\left(
\begin{array}{cc}
s_1(x,t)+i s_2(x,t) & -(s_3(-x,t)-i s_4(-x,t)) \\
s_3(x,t)+i s_4(x,t) & -(s_1(x,t)+i s_2(x,t)) \\
\end{array}\right),
\end{equation*}
where
\begin{eqnarray}\label{c2}
&&s_1(x,t)s_1(-x,t)+s_2(x,t)s_2(-x,t)+s_3(x,t)s_3(-x,t)+s_4(x,t)s_4(-x,t)=1,\nonumber\\
&&2s_1(x,t)s_2(x,t)+s_3(x,t)s_4(-x,t)-s_4(x,t)s_3(-x,t)=0,
\end{eqnarray}
and $s_j (j=1,2,3,4)$ is given by
\begin{equation*}
s_1(x,t)=\frac{s_{11}}{\Gamma},\quad s_2(x,t)=\frac{s_{12}}{\Gamma},
\quad s_3(x,t)=\frac{s_{21}}{\Gamma},\quad s_4(x,t)=\frac{s_{22}}{\Gamma},
\end{equation*}
with
\begin{align}\nonumber
\Gamma=&(a^2(x,t)+b^2(x,t))(a^2(-x,t)+b^2(-x,t))+(c^2(x,t)+d^2(x,t))(c^2(-x,t)+d^2(-x,t))\\\nonumber
&+2(a(x,t)c(x,t)+b(x,t)d(x,t))(a(-x,t)c(-x,t)+b(-x,t)d(-x,t))\\\nonumber
&+2(b(x,t)c(x,t)-a(x,t)d(x,t))(b(-x,t)c(-x,t)-a(-x,t)d(-x,t)),\\\nonumber
s_{11}=&2(a(x,t)c(x,t)+b(x,t)d(x,t))(a(-x,t)d(-x,t)-b(-x,t)c(-x,t))\\\nonumber
&+2(b(x,t)c(x,t)-a(x,t)d(x,t))(b(-x,t)d(-x,t)+a(-x,t)c(-x,t)),\\\nonumber
s_{12}=&-(a^2(x,t)+b^2(x,t))(a^2(-x,t)+b^2(-x,t))+(c^2(x,t)+d^2(x,t))(c^2(-x,t)+d^2(-x,t)),\\\nonumber
s_{21}=&-2(a^2(x,t)+b^2(x,t))(d(x,t)a(-x,t)+c(x,t)b(-x,t))\\\nonumber
&-2(c^2(x,t)+d^2(x,t))(b(x,t)c(-x,t)+a(x,t)d(-x,t)),\\\nonumber
s_{22}=&2(a^2(x,t)+b^2(x,t))(c(x,t)a(-x,t)-d(x,t)b(-x,t))\\\nonumber
&+2(c^2(x,t)+d^2(x,t))(a(x,t)c(-x,t)-b(x,t)d(-x,t)).\nonumber
\end{align}
Thus, equation \eqref{mHE} can be transformed into
\begin{align}
\frac{ds_1}{dt}=&\frac{1}{2}\left(s_3s_{3xx}(-x)-s_3(-x)s_{3xx}+s_4s_{4xx}(-x)-s_4(-x)s_{4xx}\right),\nonumber\\
\frac{ds_2}{dt}=&\frac{1}{2}\left(s_4s_{3xx}(-x)-s_3(-x)s_{4xx}+s_{3xx}s_4(-x)-s_3s_{4xx}(-x) \right),\nonumber\\
\frac{ds_3}{dt}=&s_3s_{1xx}-s_1s_{3xx}+s_2s_{4xx}-s_4s_{2xx},\nonumber\\
\frac{ds_4}{dt}=&s_4s_{1xx}-s_1s_{4xx}+s_3s_{2xx}-s_2s_{3xx}.
\end{align}
In summary, although the geometry related to the nonlocal focusing NLS and the nonlocal defocusing NLS is not very clear, we can see, from their gauge equivalence, that the properties between the nonlocal NLS equation and NLS equation have big differences.

\section{The soliton of discrete nonlocal NLS and gauge equivalence }
In this section, we will seek the soliton solution of the discrete nonlocal focusing NLS and the discrete nonlocal defocusing NLS through constructing their Darboux transformations. We will show that there is no singular point in the discrete one-soliton solution,
which is distinguished from that given in \cite{9}. We will also show that the nonlocal discrete focuing NLS and the nonlocal discrete defocuing NLS are, respectively, gauge equivalent to a discrete Heisenberg-like equation and a discrete modified Heisenberg-like equation. From the gauge equivalence, we can see that there exist big differences between the nonlocal discrete NLS equation and discrete NLS equation.\\
{\it 3.1 The soliton of discrete nonlocal focusing NLS and gauge equivalence}\\
The discrete nonlocal focusing NLS is as follows:
\begin{equation}\label{Q1}
i \frac{d Q_n}{d\tau}+Q_{n+1}+Q_{n-1}-2Q_n+ Q_n Q^{*}_{-n}(Q_{n+1}+Q_{n-1})=0,
\end{equation}
which has the discrete Lax pair
\begin{equation}\label{Q2}
E\varphi_n=M_n \varphi_n,\quad \varphi_{n,\tau}=N_n \varphi_n
\end{equation}
with
\begin{align*}
M_n=&\left(
\begin{array}{cc}
z & Q^{*}_{-n}z^{-1} \\
- Q_n z & z^{-1}  \\
\end{array}\right),\\ \nonumber
N_n=&i \left(
\begin{array}{cc}
1-z^2+z-z^{-1}- Q^{*}_{-n}Q_{n-1} & -Q^{*}_{-n}+Q^{*}_{-n-1}z^{-2} \\
- Q_n + Q_{n-1}z^2 & -1+z^{-2}+z-z^{-1}+ Q_n Q^{*}_{-n-1}  \\
\end{array}\right).
\end{align*}
We remark here that the Lax pair \eqref{Q2} is different from the one given in \cite{9}.
Equation \eqref{Q1} is $PT$ symmetric similar to the classical integrable discrete NLS.
Introduce
 $\varphi_n^{[1]}=T_n\varphi_n$, where the matrix
\begin{align}
T_n=\left(
\begin{array}{cc}
z+a_n z^{-1} & b_n z^{-1} \\
c_n z & d_n z+z^{-1}  \\
\end{array}\right),
\end{align}
with constraint condition:
\begin{equation}
b_n=-c^*_{-n}, \quad a_n=d^*_{-n}.
\end{equation}
Suppose $Q_n$ is a solution of \eqref{Q1} and $\varphi_n=(\varphi_{1,n},\varphi_{2,n})^T$ is an eigenfunction of linear problem \eqref{Q2}
with $z=z_1$. Then one can check that $\psi_n=(-\varphi^*_{2,-n},\varphi^*_{1,-n})^T$ is also the eigenfunction when $z=(z^*_1)^{-1}$.
Assume that $\det T_n(z_1)=0$, then the two column vectors in $T_n(z_1)(\varphi_n,\psi_n)$ are linear dependent. Thus we get
\begin{align}\nonumber
c_n&=\frac{(z_1^{*2}-z_1^{-2})\tau_n}{1+\tau_n \tau^*_{-n}}, \quad \quad d_n=\frac{-z_1^{*2}-z_1^{-2}\tau_n\tau^*_{-n}}{1+\tau_n \tau^*_{-n}},
\end{align}
where $\tau_n=\varphi_{2,n}/ \varphi_{1,n}$.
we have proved that the new linear problem
\begin{equation}\label{Q2new}
E\varphi^{[1]}_n=M^{[1]}_n \varphi^{[1]}_n,\quad \varphi^{[1]}_{n,\tau}=N^{[1]}_n \varphi^{[1]}_n,
\end{equation}
where
\begin{equation}\label{DT new}
M^{[1]}_n =T_{n+1} M_n T_n^{-1},\quad N^{[1]}_n=(T_{n,\tau}+T_n N_n)T_n^{-1},
\end{equation}
has the same form as the linear eigenfunction equation \eqref{Q2} except that the $Q_n, Q_{-n}^*$ in $M_n, N_n$ are replaced by $Q^{[1]}_n, Q^{*[1]}_{-n}$
in $M^{[1]}_n, N^{[1]}_n$.\\
The relation between old potential $Q_n$ and new potential $Q^{[1]}_n$ is
\begin{equation}\label{relation1}
Q^{[1]}_n=Q_n d_{n+1}-c_{n+1}.
\end{equation}
For the seed $Q_n=0$ and $z_1=\alpha+i \beta$,  the eigenfunctions are
$\varphi_{1,n}=z_1^n e^{\xi \tau}, \varphi_{2,n}=z_1^{-n} e^{\eta \tau}$ with $\xi\triangleq i(1-z_1^2+z_1-z_1^{-1}), \eta\triangleq i(-1+z_1^{-2}+z_1-z_1^{-1})$.\\
So the discrete one-soliton solution is
\begin{equation}\label{no2}
Q^{[1]}_n=\frac{z_1^{-2(n+1)}(z_1^{-2}-z_1^{*2})e^{i (z_1-z_1^{-1})^2\tau}}{1+z_1^{-2(n+1)}(z_1^*)^{2(n-1)}e^{(\eta+\eta^*-\xi-\xi^*)\tau}}.
\end{equation}
Its norm is
\begin{equation}
\left|Q^{[1]}_n\right|=\frac{(\alpha^2+\beta^2)^{-n}|(\alpha^2+\beta^2)^2-1|e^{2\alpha \beta((\alpha^2+\beta^2)^{-2}-1)\tau}}{\sqrt{\left((\alpha^2+\beta^2)^2+e^{4\alpha \beta ((\alpha^2+\beta^2)^{-2}-1)\tau}\cos\nu\right)^2+e^{8\alpha \beta((\alpha^2+\beta^2)^{-2}-1)\tau}\sin^2\nu}},
\end{equation}
where $\nu=2(n-1)\textrm{Arg}(\alpha-i\beta)-2(n+1)\textrm{Arg}(\alpha+i\beta)$.\\
If the $z_1$ is the form of polar coordinates, i.e., $z_1=\alpha e^{i\beta}$ and $\alpha\neq1, \beta\in (-\pi,\pi]$, then we get
\begin{equation}
\left|Q^{[1]}_n\right|=\frac{\alpha^{-2n}|\alpha^4-1|e^{-\left(\alpha^2-\frac{1}{\alpha^2}\right)\tau \sin(2\beta) }}{\sqrt{\alpha^8+e^{-4\left(\alpha^2-\frac{1}{\alpha^2}\right)\tau \sin(2\beta)}+2\alpha^4e^{-2\left(\alpha^2-\frac{1}{\alpha^2}\right)\tau \sin(2\beta) }\cos(4\beta n)}},
\end{equation}
Note that there is no singular point in discrete one-soliton \eqref{no2}, which is distinguished from that given in \cite{9}.  Fig. 1 gives the shape of the discrete one-soliton solution with $\alpha=\sqrt{5}/2,\beta=\arctan{1/2}$.

On the other hand, if we take the seed solution $Q_n=\rho e^{2i\rho^2 \tau+i \phi}, \quad \rho, \phi \in R$, then solving linear isospectral equation  \eqref{Q2} yields the following
eigenfunctions:
\begin{equation}
\begin{aligned}
&\varphi_{1n}=i\rho (z^{-2}-1)C^n e^{\lambda_1 \tau}+(\lambda_2-q)D^n e^{\lambda_2 \tau},\\
&\varphi_{2n}=e^{2i\rho^2 \tau+i \phi}\left(C^ne^{\lambda_1 \tau}(\lambda_1-p)+i\rho (z^{2}-1)D^n  e^{\lambda_2 \tau}\right ),
\end{aligned}
\end{equation}
where
\begin{equation}
\begin{aligned}
&\lambda_1=\frac{i(1-z^4+2z^3-2\rho^2z^2-2z)+(z^2-1)\sqrt{\Delta}}{2z^2},\\
&\lambda_2=\frac{i(1-z^4+2z^3-2\rho^2z^2-2z)-(z^2-1)\sqrt{\Delta}}{2z^2},\\
&C=\frac{z^2+1+i \sqrt{\Delta}}{2z}, \quad D=\frac{z(z^2-1-2\rho^2-i\sqrt{\Delta})}{z^2-1-i\sqrt{\Delta}},\quad \Delta=4\rho^2 z^2-(z^2-1)^2,\\
&p=i(1-z^2+z-z^{-1}-\rho^2),\quad q=i(z^{-2}+z-z^{-1}-1-\rho^2).\\
\end{aligned}
\end{equation}
So, new soliton solution is
\begin{equation}
Q^{[1]}_n=\frac{(z_1^{-2}-z_1^{*2})\tau_{n+1}}{1+\tau_{n+1} \tau^*_{-n+1}}-\rho e^{2i\rho^2 \tau+i \phi} \frac{z_1^{*2}+z_1^{-2}\tau_{n+1}\tau^*_{-n+1}}{1+\tau_{n+1} \tau^*_{-n+1}},
\end{equation}
where
\begin{equation}
\begin{aligned}
\tau_{n+1}=\frac{-\rho z^2 e^{2i\rho^2 \tau+i \phi}+\tau_n }{z^2+\rho e^{-2i\rho^2 \tau-i \phi}\tau_n }
\end{aligned}
\end{equation}
with
\begin{equation}
\begin{aligned}
&\tau_n=-e^{2i\rho^2 \tau+i \phi}\frac{i(z^2-1)+\sqrt{\Delta}+2i\rho z^2 \theta^n e^{(\lambda_2-\lambda_1)\tau}}{2i\rho+\left(i(z^2-1)+\sqrt{\Delta}\right)\theta^n e^{(\lambda_2-\lambda_1)\tau}},\\
&\theta=\frac{D}{C}=\frac{z^2(z^2-2\rho^2-1-i\sqrt{\Delta})}{z^2(1+2\rho^2)-1-i\sqrt{\Delta}}, \quad \lambda_2-\lambda_1=(z^{-2}-1)\sqrt{\Delta},\\ &\lambda_2-q=-(\lambda_1-p)=\frac{(1-z^2)\left(i(z^2-1)+\sqrt{\Delta}\right)}{2z^2}.
\end{aligned}
\end{equation}
It is interesting to note that by choosing the proper parameters, $Q_n^{[1]}$ can be rewritten as
\begin{eqnarray}
Q^{[1]}_n=\rho e^{2i\rho^2 \tau+i \phi}f(n+\gamma \tau, n-\beta\tau)
\end{eqnarray}
This means that the solution $|Q^{[1]}_n|$ is a traveling soliton solution. In the specific case, we give out the shapes
of $|Q^{[1]}_n|$.
Set $z=a+ib, a^2+b^2\neq0,\neq 1$, then we get $\Delta=-1-a^4-b^2(2+b^2+4\rho^2)+a^2(2+6 b^2+4\rho^2)+4abi(1-a^2+b^2+2\rho^2)$.
In the case of $Im\Delta=0$, i.e, $a=0$ or $b=0$ or $1-a^2+b^2+2\rho^2=0$, we discuss the properties of $Q^{[1]}_n$.\\
\textbf{Case 1:} $b=0$\\
$\bullet$ When $Re\Delta<0$, i.e., $2|\rho z|\leq |z^2-1|$, we have
\begin{equation}\label{period}
Q^{[1]}_n(\tau)=-e^{i(2\rho^2\tau+\phi)}\frac{e^{2(z^{-2}-1)\sqrt{-\Delta}i\tau}M_{1n}+e^{(z^{-2}-1)\sqrt{-\Delta}i\tau}M_{2n}+M_{3}}
{e^{2(z^{-2}-1)\sqrt{-\Delta}i\tau}M_{4n}+e^{(z^{-2}-1)\sqrt{-\Delta}i\tau}M_{5n}+M_{6}},
\end{equation}
where
\begin{equation*}
\begin{aligned}
M_{1n}=&2\theta^{2n+1}\rho^2z^2(1+z^2)(1-z^2+\sqrt{-\Delta}),\\
M_{2n}=&-\rho M_{5n}=2\theta^{n}\rho z^2(1-z^2)\left(z^2-\theta^2+(1+\theta^2)\sqrt{-\Delta}+(1+2\rho^2)(z^2\theta^2-1) \right),\\
M_{3}=&2\theta(1+z^2)\left[(1-z^6+3(1+\rho^2)z^4-(3+\rho^2)z^2+\left(\rho^2z^2-(z^2-1)^2\right)\sqrt{-\Delta}\right],\\
M_{4n}=&\theta^{2n}M_{6}, \qquad M_6=2\rho\theta z^2(1+z^2)(z^2-1+\sqrt{-\Delta}).
\end{aligned}
\end{equation*}
Note that $M_{jn}, j=1,2,...,6$ is independent of $\tau$, so $|Q^{[1]}_n(\tau)|$ is a discrete time-period soliton solution with $T=\left|\frac{2\pi}{(z^{-2}-1)\sqrt{-\Delta}}\right|$.
The discrete soliton solution $|Q_{1n}(\tau)|$ is shown in fig. 2 where the parameters $z=2, \rho=1/2$,
which is a nonzero background solution with period $T=\frac{8\pi}{3\sqrt{5}}$.
 To describe the background clearly, we give the graphs at three different space lattice points
$n=-5$, $n=0$ and $n=5$, respectively.\\

$\bullet$ When $\Delta=Re\Delta>0$, we can get a breather-like solution on space lattice $n$
\begin{equation}
Q^{[1]}_n(\tau)=-e^{i(2\rho^2\tau+\phi)}\frac{e^{2(z^{-2}-1)\sqrt{\Delta}\tau}M_{1n}+e^{(z^{-2}-1)\sqrt{\Delta}\tau}M_{2n}+M_{3n}}
{e^{2(z^{-2}-1)\sqrt{\Delta}\tau}M_{4n}+e^{(z^{-2}-1)\sqrt{\Delta}\tau}M_{5n}+M_{6n}},
\end{equation}
where
\begin{equation*}
\begin{aligned}
M_{1n}=&z^2M_{3n}^*=2\rho z^2(1+z^2)\theta^n \left(2(1+\rho^2)z^2-1-z^4-i(z^2-1) \sqrt{\Delta}\right),\\
M_{2}=&-\rho M_{5}=2\rho^2z^2(1-z^2)\left((1+z^2-i\sqrt{\Delta})\theta^{*}+(1+z^2+i\sqrt{\Delta})\theta\right),\\
M_{4n}=&z^2M_{6n}^*=4\rho^2 z^4(1+z^2)\theta^n.\\
\end{aligned}
\end{equation*}
Fig.3 shows the breather-like solution where $z=2, \rho=1$.\\
\textbf{Case 2:} For $1-a^2+b^2+2\rho^2=0$\\
In this case, we have $Re\Delta>0$. When $b\neq 0$, we find that $|Q^{[1]}_n(\tau)|$ is a two-solitons without singular
and it has local maximum value (see fig.4, where $a=2, b=\rho=1$).\\
\textbf{Case 3:} For $a=0, i.e., z=ib$\\
In this case, one can check $Re\Delta<0$. From the formula of $Q^{[1]}_n$, we see that the $|Q^{[1]}_n|$ is also a discrete time-period soliton.
Fig. 5 gives the discrete period soliton with $T=\frac{8\pi}{5\sqrt{41}}$ where $b=-1/2, \rho=1$.\\
Next we will discuss the gauge equivalence for the discrete nonlocal focusing NLS equation.
Let $S_n\triangleq G_n^{-1}\sigma_3 G_n$, where $G_n$ satisfies the linear problem
\begin{equation*}
G_{n+1}=M_n(1)G_n, \quad G_{n,\tau}=N_n(1)G_n
\end{equation*}
with the form of
\begin{align*}
G_n=\left(
\begin{array}{cc}
f_n & -g^{*}_{-n} \\
g_n & f^*_{-n}  \\
\end{array}\right),
\end{align*}
Under discrete gauge transformation
\begin{equation}
\tilde{M}_n=G_{n+1}^{-1}M_nG_n, \quad \tilde{N}_n=G_{n}^{-1}N_nG_n-G_{n}^{-1}G_{n,\tau},
\end{equation}
we obtain
\begin{align}\nonumber
\tilde{M}_n=&G_{n}^{-1}M_n^{-1}(1)M_nG_n=\frac{z+z^{-1}}{2}I+\frac{z-z^{-1}}{2}S_n, \\\nonumber
 \tilde{N}_n=&G_{n}^{-1}(N_n-N_n(1))G_n \\ \nonumber
 =&iG_n^{-1}\left(
\begin{array}{cc}
1-z^2+z-z^{-1} & (z^{-2}-1)Q^{*}_{-n-1} \\
(z^2-1)Q_{n-1} &  -1+z^{-2}+z-z^{-1}\\
\end{array}\right)G_n\\\nonumber
=&i(z-z^{-1})I+i\left(\frac{z^2+z^{-2}}{2}-1\right)G_n^{-1}\left(
\begin{array}{cc}
-1 & Q^{*}_{-n-1} \\
Q_{n-1} &  1\\
\end{array}\right)G_n\\\nonumber
&+i\frac{z^2-z^{-2}}{2}G_n^{-1}\left(
\begin{array}{cc}
-1 & -Q^{*}_{-n-1} \\
Q_{n-1} &  -1\\
\end{array}\right)G_n\\ \nonumber
=&i(z-z^{-1})I+i\left(1-\frac{z^2+z^{-2}}{2}\right)\frac{S_n+S_{n-1}}{1+\frac{1}{2}\textrm{tr}(S_nS_{n-1})}
-i\frac{z^2-z^{-2}}{2}\frac{I+S_{n-1}S_n}{1+\frac{1}{2}\textrm{tr}(S_nS_{n-1})}.
\end{align}
Here we have used the identities
\begin{align}\nonumber
&1+\frac{1}{2}\textrm{tr}(S_{n+1}S_{n})=\frac{2}{1+Q_{n}Q^{*}_{-n}},\\ \nonumber
&G_n^{-1}\left(
\begin{array}{cc}
1 & Q^{*}_{-n-1} \\
-Q_{n-1} &  1\\
\end{array}\right)G_n=G^{-1}_{n-1}G_n=\frac{I+S_{n-1}S_n}{1+\frac{1}{2}\textrm{tr}(S_nS_{n-1})}\\ \nonumber
&G_n^{-1}\left(
\begin{array}{cc}
1 & -Q^{*}_{-n-1} \\
-Q_{n-1} &  -1\\
\end{array}\right)G_n=\frac{S_n+S_{n-1}}{1+\frac{1}{2}\textrm{tr}(S_nS_{n-1})}.
\end{align}
Then by using the discrete zero curvature equation $\tilde{M}_{n,\tau}=\tilde{N}_{n+1}\tilde{M}_n-\tilde{M}_n\tilde{N}_n$ and comparing
the power of $z$, we get a discrete Heisenberg-like model
\begin{equation}\label{dH}
\frac{dS_n}{d\tau}=i\frac{[S_{n+1},S_n]}{1+\frac{1}{2}\textrm{tr}(S_{n+1}S_{n})}-i\frac{[S_{n},S_{n-1}]}{1+\frac{1}{2}\textrm{tr}(S_nS_{n-1})}.
\end{equation}
where the matrix $S_n$ is given by
\begin{equation}
S_n=G_n^{-1}\sigma_3 G_n=\frac{1}{f_nf^{*}_{-n}+g_ng^{*}_{-n}}\left(
\begin{array}{cc}
f_nf^{*}_{-n}-g_ng^{*}_{-n} & -2f^{*}_{-n}g^{*}_{-n} \\
-2f_n g_n &  g_ng^{*}_{-n}-f_nf^{*}_{-n}\\
\end{array}\right).
\end{equation}
Set $f_n=a_n+ib_n, g_n=c_n+id_n$, then $S_n$ has the form
\begin{align*}
S_n=\left(
\begin{array}{cc}
s_{1n}+i s_{2n} & s_{3(-n)}-i s_{4(-n)}  \\
s_{3n}+i s_{4n}  & -(s_{1n}+i s_{2n} ) \\
\end{array}\right),
\end{align*}
where
\begin{eqnarray}\label{c3}
&&s_{1n}s_{1(-n)}+s_{2n}s_{2(-n)}+s_{3n}s_{3(-n)}+s_{4n}s_{4(-n)}=1,\nonumber\\
&&2s_{1n}s_{2n}+s_{3(-n)}s_{4n}-s_{3n}s_{4(-n)}=0,
\end{eqnarray}
and $s_{jn}(j=1,2,3,4)$ is given by
\begin{equation*}
s_{1n}=\frac{s_n^{11}}{\Gamma_n},\quad s_{2n}=\frac{s_n^{12}}{\Gamma_n},
\quad s_{3n}=\frac{s_n^{21}}{\Gamma_n},\quad s_{4n}=\frac{s_n^{22}}{\Gamma_n},
\end{equation*}
with
\begin{align}\nonumber
\Gamma_n=&(a^2_n+b^2_n)(a^2_{-n}+b^2_{-n})+(c^2_n+d^2_n)(c^2_{-n}+d^2_{-n})
+2(a_nb_{-n}-a_{-n}b_n)(c_nd_{-n}-c_{-n}d_n)\\\nonumber
&+2(a_na_{-n}+b_nb_{-n})(c_nc_{-n}+d_nd_{-n}),\\\nonumber
s_n^{11}=&(a^2_n+b^2_n)(a^2_{-n}+b^2_{-n})-(c^2_n+d^2_n)(c^2_{-n}+d^2_{-n}),\\\nonumber
s_n^{12}=&2(a_na_{-n}+b_nb_{-n})(c_nd_{-n}-c_{-n}d_n)+2(c_nc_{-n}+d_nd_{-n})(a_{-n}b_n-a_nb_{-n}),\\\nonumber
s_n^{21}=&2(a^2_n+b^2_n)(d_nb_{-n}-c_na_{-n})+2(c^2_{n}+d^2_{n})(b_nd_{-n}-a_nc_{-n})\\\nonumber
s_n^{22}=&-2(a^2_n+b^2_n)(a_{-n}d_n+b_{-n}c_n)-2(c^2_n+d^2_n)(a_nd_{-n}+b_nc_{-n}).
\end{align}
The matrix equation \eqref{dH} can be rewritten as
\begin{equation}
\begin{aligned}
&\frac{ds_{1n}}{d\tau}=\frac{A_nE_n+B_nF_n}{\Delta_{1n}}-\frac{C_nI_n+D_nJ_n}{\Delta_{2n}},\quad
\frac{ds_{2n}}{d\tau}=\frac{A_nF_n-B_nE_n}{\Delta_{1n}}-\frac{C_nJ_n-D_nI_n}{\Delta_{2n}},\\
&\frac{ds_{3n}}{d\tau}=\frac{A_nG_n+B_nH_n}{\Delta_{1n}}-\frac{C_nK_n+D_nL_n}{\Delta_{2n}},\quad
\frac{ds_{4n}}{d\tau}=\frac{A_nH_n-B_nG_n}{\Delta_{1n}}-\frac{C_nL_n-D_nK_n}{\Delta_{2n}},
\end{aligned}
\end{equation}
where
\begin{equation}
\begin{aligned}
&\Delta_{1n}=A_n^2+B_n^2, \qquad \Delta_{2n}=C_n^2+D_n^2;\\
&A_n=1+s_{1n}s_{1(n+1)}-s_{2n}s_{2(n+1)}+\frac{1}{2}(s_{3n}s_{3(-n+1)}+s_{3(-n)}s_{3(n+1)}
+s_{4n}s_{4(-n+1)}+s_{4(-n)}s_{4(n+1)});\\
&B_n=s_{2n}s_{1(n+1)}+s_{1n}s_{2(n+1)}-\frac{1}{2}(s_{3n}s_{4(-n+1)}+s_{4(-n)}s_{3(n+1)}
-s_{4n}s_{3(-n+1)}-s_{3(-n)}s_{4(n+1)});\\
&C_n=1+s_{1n}s_{1(n-1)}-s_{2n}s_{2(n-1)}+\frac{1}{2}(s_{3(-n)}s_{3(n-1)}+s_{3n}s_{3(-n-1)}
+s_{4(n-1)}s_{4(-n)}+s_{4n}s_{4(-n-1)});\\
&D_n=s_{1n}s_{2(n-1)}+s_{2n}s_{1(n-1)}-\frac{1}{2}(s_{3n}s_{4(-n-1)}-s_{3(-n)}s_{4(n-1)}
+s_{3(n-1)}s_{4(-n)}-s_{4n}s_{3(-n-1)});\\
&E_n=s_{3n}s_{4(-n+1)}-s_{3(n+1)}s_{4(-n)}-s_{3(-n+1)}s_{4n}+s_{3(-n)}s_{4(n+1)};\\
&F_n=s_{3n}s_{3(-n+1)}-s_{3(n+1)}s_{3(-n)}+s_{4(-n+1)}s_{4n}-s_{4(-n)}s_{4(n+1)};\\
&G_n=2 (s_{2( n+1)}s_{3n}-s_{2n}s_{3(n+1)}+s_{1(n+1)}s_{4n}-s_{1n} s_{4(n+1)});\\
&H_n=2(s_{1n}s_{3(n+1)}-s_{1( n+1)}s_{3n}+s_{2(n+1)}s_{4n}-s_{2n} s_{4(n+1)});\\
&I_n=-s_{3n}s_{4(-n-1)}-s_{3(-n)}s_{4(n-1)}+s_{3(n-1)}s_{4(-n)}+s_{3(-n-1)}s_{4n};\\
&J_n=s_{3(-n)}s_{3(n-1)}-s_{3(n)}s_{3(-n-1)}+s_{4(n-1)}s_{4(-n)}-s_{4(-n-1)}s_{4n};\\
&K_n=2(s_{2n}s_{3(n-1)}-s_{2(n-1)}s_{3n}+s_{1n}s_{4(n-1)}-s_{1(n-1)}s_{4n});\\
&L_n=2(s_{1(n-1)}s_{3n}-s_{1n}s_{3(n-1)}+s_{2n}s_{4(n-1)}-s_{2(n-1)}s_{4n}).
\end{aligned}
\end{equation}
We remark here that if set $f^{*}_{-n}=f^{*}_{n}, g^{*}_{-n}=g^{*}_{n}$ and $f_n=a_n+ib_n, g_n=c_n+id_n$, then we can show that the matrix $S_n$ possesses the form
\begin{equation}
S_n=\left(
\begin{array}{cc}
s_{1n}& s_{2n}-is_{3n}\\
 s_{2n}+is_{3n} &  -s_{1n}\\
\end{array}\right),
\end{equation}
where the vector $\textbf{S}_n$=$(s_{1n},s_{2n},s_{3n})^T\in S^2$ in $R^3$, and $s_{jn}, j=1,2,3$ is given by
\begin{eqnarray*}
s_{1n}=\frac{|f_n|^2-|g_n|^2}{|f_n|^2+|g_n|^2},\qquad
s_{2n}=\frac{2(b_nd_n-a_nc_n)}{|f_n|^2+|g_n|^2},\qquad
s_{3n}=\frac{-2(b_nc_n+a_nd_n)}{|f_n|^2+|g_n|^2},
\end{eqnarray*}
and equation \eqref{dH} reduces to
\begin{align}
\dot{\textbf{S}}_{n}=2
\frac{\textbf{S}_{n+1}\times\textbf{S}_{n}}{1+\textbf{S}_{n+1}\cdot \textbf{S}_{n}}-2\frac{\textbf{ S}_{n}\times\textbf{S}_{n-1}}{1+\textbf{S}_{n}\cdot \textbf{S}_{n-1}}.
\end{align}
This means that the equation \eqref{dH} is a discrete Heisenberg equation under the condition $f^{*}_{-n}=f^{*}_{n}, g^{*}_{-n}=g^{*}_{n}$ which agrees with a well-known fact that
the discrete focusing NLS is gauge equivalent to the discrete Heisenberg equation \cite{15}.\\
{\it 3.2  The soliton of discrete nonlocal defocusing NLS and gauge equivalence}\\
For the discrete nonlocal defocusing NLS
\begin{equation}\label{fu1}
i \frac{d{Q}_n}{d\tau}+ Q_{n+1}+Q_{n-1}-2Q_n- Q_n Q^{*}_{-n}(Q_{n+1}+Q_{n-1})=0,
\end{equation}
its Lax pair is:
\begin{equation}\label{fu2}
E\varphi_n=M_n \varphi_n,\quad \varphi_{n,\tau}=N_n \varphi_n
\end{equation}
with
\begin{align*}
M_n=&\left(
\begin{array}{cc}
z & Q^{*}_{-n}z^{-1} \\
Q_n z & z^{-1}  \\
\end{array}\right),\\ \nonumber
N_n=&i \left(
\begin{array}{cc}
1-z^2+z-z^{-1}+Q^{*}_{-n}Q_{n-1} & -Q^{*}_{-n}+Q^{*}_{-n-1}z^{-2} \\
 Q_n - Q_{n-1}z^2 & -1+z^{-2}+z-z^{-1}- Q_n Q^{*}_{-n-1}  \\
\end{array}\right).
\end{align*}
We also remark here that the Lax pair \eqref{fu2} is different from the one given in \cite{9}.
We have obtained the following Darboux transformation for discrete nonlocal defocusing NLS equation \eqref{fu1}:
\begin{align}
\varphi_n^{[1]}=\left(
\begin{array}{cc}
z+a_n z^{-1} & b_n z^{-1} \\
c_n z & d_n z+z^{-1}  \\
\end{array}\right)\varphi_n,
\end{align}
with constraint condition:
\begin{equation}
b_n=c^*_{-n}, \quad a_n=d^*_{-n}.
\end{equation}
Further, we have
\begin{align}\nonumber
c_n&=\frac{(z_1^{*2}-z_1^{-2})\tau_n}{1-\tau_n \tau^*_{-n}}, \quad \quad d_n=\frac{-z_1^{*2}-z_1^{-2}\tau_n\tau^*_{-n}}{1-\tau_n \tau^*_{-n}},
\end{align}
and the relation between old potential $Q_n$ and new potential $Q^{[1]}_n$ can be written as
\begin{equation}\label{relation2}
Q^{[1]}_n=c_{n+1}+Q_n d_{n+1}.
\end{equation}
For seed solution $Q_n=0$, one can get discrete 1-soliton solution for discrete nonlocal defocusing NLS
\begin{equation}\label{s1}
Q^{[1]}_n=\frac{z_1^{-2(n+1)}(z_1^{*2}-z_1^{-2})e^{i (z_1-z_1^{-1})^2\tau}}{1-z_1^{-2(n+1)}(z_1^*)^{2(n-1)}e^{(\eta+\eta^*-\xi-\xi^*)\tau}}.
\end{equation}
where $\xi\triangleq i(1-z_1^2+z_1-z_1^{-1}), \eta\triangleq i(-1+z_1^{-2}+z_1-z_1^{-1})$.
Set $z_1=\alpha e^{i\beta}$ and $\alpha\neq1, \beta\in (-\pi,\pi]$, then we get
\begin{equation}
\left|Q^{[1]}_n\right|=\frac{\alpha^{-2n}|\alpha^4-1|e^{-\left(\alpha^2-\frac{1}{\alpha^2}\right)\tau \sin(2\beta) }}{\sqrt{\alpha^8+e^{-4\left(\alpha^2-\frac{1}{\alpha^2}\right)\tau \sin(2\beta)}-2\alpha^4e^{-2\left(\alpha^2-\frac{1}{\alpha^2}\right)\tau \sin(2\beta) }\cos(4\beta n)}}.
\end{equation}
Its shape is given in Fig.6 with $\alpha=\sqrt{5}/2,\beta=\arctan{1/2}$. The solution \eqref{s1} differs from one of the classical discrete defocusing NLS.
Taking the seed solution
$$Q_n=\rho e^{-2i\rho^2 \tau+i \phi}, \quad \rho, \phi \in R$$
and solving equation \eqref{fu2} give the following
eigenfunctions:
\begin{equation}
\begin{aligned}
&\varphi_{1n}=i\rho (z^{-2}-1)C^n e^{\lambda_1 \tau}+(\lambda_2-q)D^n e^{\lambda_2 \tau},\\
&\varphi_{2n}=e^{-2i\rho^2 \tau+i \phi}\left(C^ne^{\lambda_1 \tau}(\lambda_1-p)+i\rho (1-z^{2})D^n  e^{\lambda_2 \tau}\right ),
\end{aligned}
\end{equation}
where
\begin{equation}
\begin{aligned}
&\lambda_1=\frac{i(1-z^4+2z^3-2\rho^2z^2-2z)+(z^2-1)\sqrt{\Delta}}{2z^2},\\
&\lambda_2=\frac{i(1-z^4+2z^3-2\rho^2z^2-2z)-(z^2-1)\sqrt{\Delta}}{2z^2},\\
&C=\frac{z^2+1+i \sqrt{\Delta}}{2z}, \quad D=\frac{z(z^2-1+2\rho^2-i\sqrt{\Delta})}{z^2-1-i\sqrt{\Delta}},\quad \Delta=-4\rho^2 z^2-(z^2-1)^2,\\
&p=i(1-z^2+z-z^{-1}+\rho^2),\quad q=i(z^{-2}+z-z^{-1}-1+\rho^2).\\
\end{aligned}
\end{equation}
We thus obtain  new soliton solution
\begin{equation}
Q^{[1]}_n=\frac{(z_1^{*2}-z_1^{-2})\tau_{n+1}}{1-\tau_{n+1} \tau^*_{-n+1}}+\rho e^{-2i\rho^2 \tau+i \phi} \frac{-z_1^{*2}+z_1^{-2}\tau_{n+1}\tau^*_{-n+1}}{1-\tau_{n+1} \tau^*_{-n+1}},
\end{equation}
with
\begin{equation}
\begin{aligned}
&\tau_{n+1}=-e^{-2i\rho^2 \tau+i \phi}\frac{i(z^2-1)+\sqrt{\Delta}-2i\rho z^2 \theta^{n+1} e^{(\lambda_2-\lambda_1)\tau}}{2i\rho+\left(i(z^2-1)+\sqrt{\Delta}\right)\theta^{n+1} e^{(\lambda_2-\lambda_1)\tau}};\\
&\theta=\frac{D}{C}=\frac{z^2(z^2+2\rho^2-1-i\sqrt{\Delta})}{z^2(1-2\rho^2)-1-i\sqrt{\Delta}}, \quad \lambda_2-\lambda_1=(z^{-2}-1)\sqrt{\Delta},\\ &\lambda_2-q=-(\lambda_1-p)=\frac{(1-z^2)\left(i(z^2-1)+\sqrt{\Delta}\right)}{2z^2}.
\end{aligned}
\end{equation}
Let us discuss the properties of $Q^{[1]}_n$.
Set $z=a+ib, a^2+b^2\neq0,\neq 1$, then we have $\Delta=-1-a^4-b^2(2+b^2-4\rho^2)+a^2(2+6 b^2-4\rho^2)+4abi(1-a^2+b^2-2\rho^2)$.
Suppose $Im\Delta=0$, i.e, $a=0$ or $b=0$ or $1-a^2+b^2-2\rho^2=0$.\\
\textbf{Case 1:} For $b=0$\\
In this case, due to $Re\Delta<0$,  we find that $|Q^{[1]}_n(\tau)|$ is a even discrete soliton on time $\tau$ when $|Q^{[1]}_n(\tau)|$ does not
exist singularity for some parameters. Fig.7 gives the discrete soliton $|Q^{[1]}_n(\tau)|$ where $b=2, \rho=-0.5,\phi=0$.
Note that $|Q^{[1]}_n(\tau)|$ is related to $\phi$. When $\phi\neq 0$,  $|Q^{[1]}_n(\tau)|$ has a shift along $\tau$.
\begin{equation}
Q^{[1]}_n(\tau)=\rho e^{-i2\rho^2\tau+i\phi}\frac{e^{2(z^{-2}-1)\sqrt{-\Delta}i\tau}M_{1n}(\tau)+e^{(z^{-2}-1)\sqrt{-\Delta}i\tau}M_{2n}(\tau)+M_{3n}(\tau)}
{e^{2(z^{-2}-1)\sqrt{-\Delta}i\tau}M_{4n}+e^{(z^{-2}-1)\sqrt{-\Delta}i\tau}M_{5n}+M_{6}},
\end{equation}
where
\begin{equation*}
\begin{aligned}
M_{1n}(\tau)=&2\rho z^2\theta^{2n+1}\left[A\left(e^{2i\rho^2 \tau-i\phi}z^2-\rho (1+z^2)\right)-2\rho\right] ,\\
M_{2n}(\tau)=&\theta^{n} \left[\left(2 A(1-\theta^2z^2)+A^2(1-\theta^2z^4)+4z^4\rho^2(\theta^2-1)\right)\rho+z^4e^{2i\rho^2 \tau-i\phi}(A^2\theta^2+4\rho^2) \right],\\
M_{3n}(\tau)=&A \theta(A+2 \rho z^4e^{2i\rho^2 \tau-i\phi}-2z^2\rho^2-z^4\rho^2),\\
M_{4n}=&\theta^{2n}M_6=2A\theta^{1+2n} z^2(1+z^2)\rho^2,\\
M_{5n}=&\rho z^2\theta^n\left( A^2(\theta^2-1)-4\rho^2(z^4\theta^2-1)\right),
\end{aligned}
\end{equation*}
with $A=z^2-1+\sqrt{-\Delta}.$\\
\textbf{Case 2:} For $a=0$\\
When $Re\Delta>0$, by choosing proper parameters, we get a discrete breather-like soliton (see fig.8)
\begin{equation}
Q^{[1]}_n(\tau)=e^{-i(2\rho^2\tau+\phi)}\frac{e^{2(z^{-2}-1)\sqrt{\Delta}\tau}M_{1n}+e^{(z^{-2}-1)\sqrt{\Delta}\tau}M_{2n}+M_{3n}}
{e^{2(z^{-2}-1)\sqrt{\Delta}\tau}M_{4n}+e^{(z^{-2}-1)\sqrt{\Delta}\tau}M_{5n}+M_{6n}},
\end{equation}
where
\begin{equation*}
\begin{aligned}
M_{1n}=&\rho z^2\theta^n \left[4z^2\rho^2-\left(Az^2-2i(z^4-1)\right)A^*\right],\\
M_{2n}=&\left(2iAz^2\rho^2+(A-Az^4-2iz^4\rho^2)A^*\right)\theta^*+2z^2\rho^2\theta(-2+iAz^2+2z^4-iA^*),\\ M_{3n}=&\rho\theta^{*n}\left(2iA(z^4-1)-4z^4\rho^2+AA^*\right);\\
M_{4n}=&\theta^n z^2(AA^*-4z^4\rho^2),\qquad M_{6n}=z^2\theta^*(4\rho^2-AA^*),\\
M_{5n}=&2iz^2\rho\left((A^*-Az^2)\theta^*+\theta(z^2\theta^*-A)\right),\\
\end{aligned}
\end{equation*}
with $A=i(z^2-1)+\sqrt{\Delta}.$\\
\textbf{Case 3:} $\Delta\neq0$.
In this case, we can get a discrete 2-soliton solution with nonlocal maximum value when $a=0.5, b=1,$ and $\rho=2$ (see fig. 9).\\
Next, we will discuss gauge equivalence of discrete nonlocal defocusing NLS. Set $S_n\triangleq -iG_n^{-1}\sigma_3 G_n$, where $G_n$ satisfies the linear problem
\begin{equation*}
G_{n+1}=M_n(1)G_n, \quad G_{n,\tau}=N_n(1)G_n
\end{equation*}
with the form of
\begin{align*}
G_n=\left(
\begin{array}{cc}
f_n & g_n  \\
f^*_{-n}& -g^{*}_{-n}  \\
\end{array}\right).
\end{align*}
Under discrete gauge transformation
\begin{equation}
\tilde{M}_n=G_{n+1}^{-1}M_nG_n, \quad \tilde{N}_n=G_{n}^{-1}N_nG_n-G_{n}^{-1}G_{n,\tau},
\end{equation}
we obtain
\begin{align}\nonumber
\tilde{M}_n=&-iG_{n}^{-1}M_n^{-1}(1)M_nG_n=\frac{z+z^{-1}}{2}I+i\frac{z-z^{-1}}{2}S_n, \\\nonumber
 \tilde{N}_n=&G_{n}^{-1}(N_n-N_n(1))G_n \\ \nonumber
 =&iG_n^{-1}\left(
\begin{array}{cc}
1-z^2+z-z^{-1} & (z^{-2}-1)Q^{*}_{-n-1} \\
(1-z^2)Q_{n-1} &  -1+z^{-2}+z-z^{-1}\\
\end{array}\right)G_n\\\nonumber
=&i(z-z^{-1})I+i\left(\frac{z^2+z^{-2}}{2}-1\right)G_n^{-1}\left(
\begin{array}{cc}
-1 & Q^{*}_{-n-1} \\
-Q_{n-1} &  1\\
\end{array}\right)G_n\\\nonumber
&+i\frac{z^2-z^{-2}}{2}G_n^{-1}\left(
\begin{array}{cc}
-1 & -Q^{*}_{-n-1} \\
-Q_{n-1} &  -1\\
\end{array}\right)G_n\\ \nonumber
=&i(z-z^{-1})I+\frac{z^2+z^{-2}-2}{2}\frac{S_n+S_{n-1}}{1-\frac{1}{2}\textrm{tr}(S_nS_{n-1})}
-i\frac{z^2-z^{-2}}{2}\frac{I-S_{n-1}S_n}{1-\frac{1}{2}\textrm{tr}(S_nS_{n-1})}.
\end{align}
Here we have used the identities
\begin{align}\nonumber
&1-\frac{1}{2}\textrm{tr}(S_{n+1}S_{n})=\frac{2}{1-Q_{n}Q^{*}_{-n}},\\ \nonumber
&G_n^{-1}\left(
\begin{array}{cc}
1 & Q^{*}_{-n-1} \\
Q_{n-1} &  1\\
\end{array}\right)G_n=G^{-1}_{n-1}G_n=\frac{I-S_{n-1}S_n}{1-\frac{1}{2}\textrm{tr}(S_nS_{n-1})}\\ \nonumber
&G_n^{-1}\left(
\begin{array}{cc}
1 & -Q^{*}_{-n-1} \\
Q_{n-1} &  -1\\
\end{array}\right)G_n=i\frac{S_n+S_{n-1}}{1-\frac{1}{2}\textrm{tr}(S_nS_{n-1})}.
\end{align}
Then by using the discrete zero curvature equation $\tilde{M}_{n,\tau}=\tilde{N}_{n+1}\tilde{M}_n-\tilde{M}_n\tilde{N}_n$ and comparing
the power of $z$, we get a discrete modified Heisenberg-like model
\begin{equation}\label{dmH2}
\frac{dS_n}{dt}=\frac{[S_{n},S_{n-1}]}{1-\frac{1}{2}\textrm{tr}(S_nS_{n-1})}-\frac{[S_{n+1},S_n]}{1-\frac{1}{2}\textrm{tr}(S_{n+1}S_{n})},
\end{equation}
where the matrix $S_n$ is given by
\begin{equation}
S_n=-iG_n^{-1}\sigma_3 G_n=\frac{i}{f_ng^{*}_{-n}+g_nf^{*}_{-n}}\left(
\begin{array}{cc}
g_n f^{*}_{-n}-f_n g^{*}_{-n} & -2g_{n}g^{*}_{-n} \\
-2f_n f^{*}_{-n} &  f_ng^{*}_{-n}-g_nf^{*}_{-n}\\
\end{array}\right).
\end{equation}
Let us further investigate the structure of the matrix $S_n$. Set $f_n=a_n+ib_n, g_n=c_n+id_n$, then $S_n$ has the form
\begin{align*}
S_n=\left(
\begin{array}{cc}
s_{1n}+i s_{2n} & s_{5n}+i s_{6n}  \\
s_{3n}+i s_{4n}  & -(s_{1n}+i s_{2n} ) \\
\end{array}\right),
\end{align*}
where
\begin{eqnarray}\label{c4}
&&s_{1n}s_{1(-n)}+s_{2n}s_{2(-n)}-s_{3n}s_{5n}+s_{4n}s_{6n}=1, \nonumber\\
&&2s_{1n}s_{2n}+s_{3n}s_{6n}+s_{4n}s_{5n}=0,
\end{eqnarray}
and $s_{jn}
(j=1,2,3,4,5,6) $ is given by
\begin{equation*}
\begin{aligned}
&s_{1n}=\frac{s_n^{11}}{\Gamma_n},\quad s_{2n}=\frac{s_n^{12}}{\Gamma_n},\quad s_{3n}=\frac{s_n^{21}}{\Gamma_n},\\
&s_{4n}=\frac{s_n^{22}}{\Gamma_n},\quad s_{5n}=\frac{s_n^{31}}{\Gamma_n},\quad s_{6n}=\frac{s_n^{32}}{\Gamma_n},
\end{aligned}
\end{equation*}
with
\begin{equation*}
\begin{aligned}\nonumber
\Gamma_n=&(a^2_n+b^2_n)(c^2_{-n}+d^2_{-n})+(a^2_{-n}+b^2_{-n})(c^2_n+d^2_n)
+2(a_nb_{-n}+a_{-n}b_n)(c_nd_{-n}+c_{-n}d_n)\\
&+2(a_na_{-n}-b_nb_{-n})(c_nc_{-n}-d_nd_{-n}),\\
s_n^{11}=&2(b_nb_{-n}-a_na_{-n})(c_nd_{-n}+c_{-n}d_n)+2(c_nc_{-n}-d_nd_{-n})(a_{-n}b_n+a_nb_{-n}),\\
s_n^{12}=&(a^2_{-n}+b^2_{-n})(c^2_n+d^2_n)-(a^2_n+b^2_n)(c^2_{-n}+d^2_{-n}),\\
s_n^{21}=&2(a^2_n+b^2_n)(b_{-n}c_{-n}-a_{-n}d_{-n})+2(a^2_{-n}+b^2_{-n})(a_nd_{n}-b_nc_{n}),\\
s_n^{22}=&2(a^2_n+b^2_n)(a_{-n}c_{-n}+b_{-n}d_{-n})+2(a^2_{-n}+b^2_{-n})(a_nc_{n}+b_nd_{n}),\\
s_n^{31}=&2(c^2_n+d^2_n)(a_{-n}d_{-n}-b_{-n}c_{-n})+2(c^2_{-n}+d^2_{-n})(b_nc_{n}-a_nd_{n}),\\
s_n^{32}=&2(c^2_n+d^2_n)(a_{-n}c_{-n}+b_{-n}d_{-n})+2(c^2_{-n}+d^2_{-n})(a_nc_{n}+b_nd_{n}).
\end{aligned}
\end{equation*}
\begin{equation*}
\begin{aligned}
&\frac{ds_{1n}}{d\tau}=\frac{C_nI_n+D_nJ_n}{\Delta_{1n}}-\frac{A_nE_n+B_nF_n}{\Delta_{2n}},\quad
\frac{ds_{2n}}{d\tau}=\frac{C_nJ_n-D_nI_n}{\Delta_{1n}}-\frac{A_nF_n-B_nE_n}{\Delta_{2n}},\\
&\frac{ds_{3n}}{d\tau}=\frac{C_nK_n+D_nL_n}{\Delta_{1n}}-\frac{A_nP_n+B_nH_n}{\Delta_{2n}},\quad
\frac{ds_{4n}}{d\tau}=\frac{C_nL_n-D_nK_n}{\Delta_{1n}}-\frac{A_nH_n-B_nP_n}{\Delta_{2n}},\\
&\frac{ds_{5n}}{d\tau}=\frac{C_nX_n+D_nY_n}{\Delta_{1n}}-\frac{A_nR_n+B_nW_n}{\Delta_{2n}},\quad
\frac{ds_{6n}}{d\tau}=\frac{C_nY_n-D_nX_n}{\Delta_{1n}}-\frac{A_nW_n-B_nR_n}{\Delta_{2n}},
\end{aligned}
\end{equation*}
with
\begin{equation*}
\begin{aligned}
\Delta_{1n}&=C_n^2+D_n^2, \qquad \Delta_{2n}=A_n^2+B_n^2;\\
A_n=&\frac{1}{2}\left(2-s_{1n}(s_{1(n+1)}-s_{1(n-1)})+s_{2n}(s_{2(n+1)}-s_{2(n-1)})-s_{3(n-1)}s_{5n}\right.\\
&\left.-s_{3n}s_{5(n+1)}+s_{4(n-1)}s_{6n}+s_{4n}s_{6(n+1)}\right);\\
B_n=&\frac{-1}{2}(s_{1n}s_{2(n-1)}+s_{2n}s_{1(n-1)}+s_{2n}s_{1(n+1)}+s_{1n}s_{2(n+1)}+s_{5n}s_{4(n-1)}\\
&+s_{4n}s_{5(n+1)}+s_{6n}s_{3(n-1)}+s_{3n}s_{6(n+1)});\\
C_n=&\frac{1}{2}(2-2s_{1n}s_{1(n-1)}+2s_{2n}s_{2(n-1)}-s_{3n}s_{5(n-1)}-s_{5n}s_{3(n-1)}
+s_{4n}s_{6(n-1)}+s_{6n}s_{4(n-1)});\\
D_n=&\frac{-1}{2}(2s_{1n}s_{2(n-1)}+2s_{2n}s_{1(n-1)}+s_{4n}s_{5(n-1)}+s_{5n}s_{4(n-1)}
+s_{3n}s_{6(n-1)}+s_{6n}s_{3(n-1)});\\
E_n=&s_{3n}s_{5(n+1)}-s_{5n}s_{3(n+1)}+s_{6n}s_{4(n+1)}-s_{4n}s_{6(n+1)};\\
F_n=&s_{4n}s_{5(n+1)}-s_{5n}s_{4(n+1)}-s_{6n}s_{3(n+1)}-s_{3n}s_{6(n+1)};\\
P_n=&2(s_{1n}s_{3(n+1)}-s_{3n}s_{1(n+1)}+s_{4n}s_{2(n+1)}-s_{2n}s_{4(n+1)});\\
H_n=&2(s_{2n}s_{3(n+1)}-s_{3n}s_{2( n+1)}-s_{4n}s_{1(n+1)}+s_{1n}s_{4(n+1)});\\
I_n=&s_{5n}s_{3(n-1)}-s_{3n}s_{5(n-1)}+s_{4n}s_{6(n-1)}-s_{6n}s_{4(n-1)};\\
J_n=&s_{5n}s_{4(n-1)}-s_{4n}s_{5(n-1)}-s_{3n}s_{6(n-1)}+s_{6n}s_{3(-n-1)};\\
K_n=&2(s_{3n}s_{1(n-1)}-s_{1n}s_{3(n-1)}+s_{2n}s_{4(n-1)}-s_{4n}s_{2(n-1)});\\
L_n=&2(s_{3n}s_{2(n-1)}-s_{2n}s_{3(n-1)}-s_{1n}s_{4(n-1)}+s_{4n}s_{1(n-1)});\\
R_n=&2(s_{5n}s_{1(n+1)}-s_{1n}s_{5(n+1)}-s_{6n}s_{2(n+1)}+s_{2n}s_{6(n+1)});\\
W_n=&2(s_{5n}s_{2(n+1)}-s_{2n}s_{5(n+1)}+s_{6n}s_{1(n+1)}-s_{1n}s_{6(n+1)});\\
X_n=&2(s_{1n}s_{5(n-1)}-s_{5n}s_{1(n-1)}-s_{2n}s_{6(n-1)}+s_{6n}s_{2(n-1)});\\
Y_n=&2(s_{2n}s_{5(n-1)}-s_{5n}s_{2(n-1)}+s_{1n}s_{6(n-1)}-s_{6n}s_{1(n-1)}).
\end{aligned}
\end{equation*}
It is interesting to note that if set $f^{*}_{-n}=f^{*}_{n}, g^{*}_{-n}=g^{*}_{n}$ and $f_n=a_n+ib_n, g_n=c_n+id_n$, then we can show that the matrix $S_n$ possesses the form
\begin{equation}
S_n=\left(
\begin{array}{cc}
s_{1n}& i(s_{3n}-s_{2n})\\
i( s_{3n}+s_{2n}) &  -s_{1n}\\
\end{array}\right),
\end{equation}
where the vector $\textbf{S}_n$=$(s_{1n},s_{2n},s_{3n})^T\in H^2$ in $R^{2+1}$, i.e., $s_{1n}^2+s_{2n}^2-s_{3n}^2=-1, $ and $s_{jn}, j=1,2,3$ is given by
\begin{eqnarray*}
s_{1n}=\frac{a_nd_n-b_nc_n}{a_nc_n+b_nd_n},\qquad
s_{2n}=\frac{|g_n|^2-|f_n|^2}{2(a_nc_n+b_nd_n)},\qquad
s_{3n}=\frac{-|f_n|^2-|g_n|^2}{2(a_nc_n+b_nd_n)}.
\end{eqnarray*}
Thus, equation \eqref{dmH2} leads to
\begin{align}
\dot{\textbf{S}}_{n}=2
\frac{\textbf{ S}_{n}\dot{\times}\textbf{S}_{n-1}}{1-\textbf{S}_{n}\cdot \textbf{S}_{n-1}}-2
\frac{\textbf{S}_{n+1}\dot{\times}\textbf{S}_{n}}{1-\textbf{S}_{n+1}\cdot \textbf{S}_{n}}.
\end{align}
This means that the equation \eqref{dmH2} is a discrete modified Heisenberg equation under the condition $f^{*}_{-n}=f^{*}_{n}, g^{*}_{-n}=g^{*}_{n}$ which agrees with a well-known fact that the discrete defocusing NLS is gauge equivalent to the discrete modified Heisenberg equation \cite{16}.\\


\section{Conclusion and Discussion}
In this paper, we have shown that, under the gauge transformations, the nonlocal focusing NLS (the nonlocal defocusing NLS) and its discrete version, discrete nonlocal focusing NLS (discrete nonlocal defocusing NLS) are, respectively, gauge equivalent to a Heisenberg-like equation (modified Heisenberg-like equation) and a discrete Heisenberg-like equation (discrete modified Heisenberg-like equation). From the gauge equivalence, we can see that the properties between the nonlocal NLS and its discrete version and NLS and discrete NLS have great differences. We have also obtained the discrete soliton solutions for the discrete nonlocal NLS through constructing the Darboux transformation. These discrete soliton solutions are different from ones obtained by using the scattering transformation. We should point out here that
geometric interpretation for the nonlocal NLS and its discrete version is not very clear at the moment. How to understand the meaning of equations \eqref{c1}, \eqref{c2},\eqref{c3}, and \eqref{c4}? This problem is worth a further investigation in the future.\\

\vskip 16pt \noindent {\bf
Acknowledgements} \vskip 12pt

The work of ZNZ is supported by the National Natural Science
Foundation of China under grant 11271254, that of ZNZ
in part by the Ministry of Economy and Competitiveness of Spain under
contract MTM2012-37070. \\

\small{

}

\end{document}